# Observation of Fluctuation Spin Hall Effect in Antiferromagnet

Chi Fang[†,1,2], Caihua Wan[†,1,7], Xiaoyue Zhang[3,5], Satoshi Okamoto[4], Tianyi Ma[1,3], Jianying Qin[1,3], Xiao Wang[1], Chenyang Guo[1], Jing Dong[1], Guoqiang Yu[1,7], Zhenchao Wen[6], Ning Tang[5], Stuart S. P. Parkin[2], Naoto Nagaosa[9], Yuan Lu[3*], Xiufeng Han[1,7,8*]

[1]*Beijing National Laboratory for Condensed Matter Physics, Institute of Physics, University of Chinese Academy of Sciences, Chinese Academy of Sciences, Beijing 100190, China*

[2]*Max Planck Institute of Microstructure Physics, Halle (Saale) 06120, Germany*

[3]*Université de Lorraine, CNRS, Institut Jean Lamour, UMR 7198, campus ARTEM, 2 Allée André Guinier, 54011 Nancy, France*

[4]*Materials Science and Technology Division, Oak Ridge National Laboratory, Oak Ridge, TN 37831, USA*

[5]*State Key Laboratory of Artificial Microstructure and Mesoscopic Physics, School of Physics, Peking University, Beijing 100871, China*

[6]*National Institute for Materials Science NIMS, Tsukuba, Ibaraki 3050047, Japan*

[7]*Songshan Lake Materials Laboratory, Dongguan, Guangdong 523808, China*

[8]*Center of Materials Science and Optoelectronics Engineering, University of Chinese Academy of Sciences, Beijing 100049, China*

[9]*RIKEN Center for Emergent Matter Science (CEMS), Wako, 351-0198, Japan.*

†These two authors contributed equally to this work.

*Corresponding Authors: yuan.lu@univ-lorraine.fr, xfhan@iphy.ac.cn,

**Abstract**

The spin Hall effect (SHE) can generate a pure spin current by an electric current, which is promisingly used to electrically control magnetization. To reduce power consumption of this control, a giant spin Hall angle (SHA) in the SHE is desired in low-resistivity systems for practical applications. Here, critical spin fluctuation near the antiferromagnetic (AFM) phase-transition is proved as an effective mechanism to create an additional part of SHE, named as fluctuation spin Hall effect (FSHE). This FSHE enhances the SHA due to the AFM spin fluctuation between conduction electrons and local spins. We detect the FSHE with the inverse and direct spin Hall effect (ISHE and DSHE) set-up and their temperature ($T$) dependences in the Cr/MgO/Fe magnetic tunnel junctions (MTJs). The SHA is significantly enhanced when temperature is approached to the Néel temperature ($T_N$) and has a peak value of -0.34 at 200 K near $T_N$. This value is higher than the room-temperature value by 240% and comparable to that of heavy metals Ta and W. Furthermore, the spin Hall resistivity of Cr well fits the modeled $T$-dependence when $T$ approaches $T_N$ from low temperatures, implying the AFM spin fluctuation nature of strong SHA enhancement. Thus, this study demonstrates the critical spin fluctuation as a prospective way of increasing SHA and enriches the AFM material candidates for spin-orbitronic devices.

The spin Hall effect utilizes spin-orbit coupling to convert a longitudinal charge current $j_c$ into a transverse pure spin current $j_s$. $j_s$ at the interface can be further absorbed by a magnetic thin film, exert a spin-orbit torque (SOT) on the magnet and finally drive its switching or oscillation. Owing to the electrical controllability on magnetization, the SHE shows promising applications in magnetic random-access memory (MRAM)[1-3], programmable logic devices[4,5] and microwave nano-oscillators[6].

SHA ($\theta_{SH}$) defined as the ratio between $j_s$ and $j_c$, or alternatively, the ratio between spin Hall conductivity $\sigma_{SH}$ and longitudinal conductivity $\sigma$. Companied by the factor of high $\sigma$, $\theta_{SH}$ is the key factor to reduce writing power and achieve high efficiency of SOT devices[7]. Thus, it is appealing to develop materials and/or explore physics to improve $\theta_{SH}$ in low-resistive systems[8,9].

The microscopic mechanisms behind the SHE have been extensively investigated. They can be categorized into three pictures, the intrinsic one due to the nontrivial Berry curvatures of electronic band structures and the other two extrinsic ones, the side-jump(sj) and skew-scattering(ss) mechanisms[8]. Inspired by the mechanisms, several effective means have been developed to enhance the SHA, such as electronic structure engineering[10], doping[11], interface decoration[12] and superlattice stacking[13]. For example, because of the intrinsic mechanism, the heavy metals Ta[14,15], Pt[16-18], W[19] with large atomic numbers[18] or some topological insulators with symmetry-protected surface states[20,21] are found to own large $\theta_{SH}$. The extrinsic mechanisms are taken into account when the spin-orbit-coupling (SOC) related scatterings between delocalized electrons and local spins are considered. It is reported that doping heavy-metal dopants into a small-SHA matrix or reverse could remarkably enhance $\theta_{SH}$ of the matrix material, such as Ir-doped Cu[22] and Pt-doped Au[7].

Significantly different from the SHE in normal metals, the FSHE contribution we reported here only involves the two extrinsic mechanisms excluding the intrinsic one since thermally-activated fluctuation of spin lattice can only contribute to the former, which can be deemed as an evidence to show the impact of the extrinsic mechanism to the overall SHE. Recently, critical spin fluctuation in magnetically-ordered systems at their magnetic phase transition temperatures $T_C$ was reported to nontrivially result in an elevated $\theta_{SH}$ by intensifying interaction of delocalized electrons with local spins. For ferromagnets, Wei et al.[23] found an ISHE anomaly near $T_C$ in the weak ferromagnet NiPd alloys. Ou et al.[24] observed an enhanced $\theta_{SH} \sim 0.34$ in the $Fe_xPt_{1-x}$ alloy at its $T_C$. Wu et al.[25] observed a $\theta_{SH} \sim 0.46$ in $Ni_xCu_{1-x}$ alloys due to spin fluctuation around $T_C$. For antiferromagnets, Saglam et al.[26] evidenced that spin fluctuation of the antiferromagnetic FeMn at its Néel temperature increased $\theta_{SH}$ of the $Ni_{80}Fe_{20}$/FeMn/W trilayer in a spin-pumping experiment. These ferromagnetic and antiferromagnetic systems have clearly hinted the contribution of spin fluctuations to the enhanced SHE phenomenally; however, all the above systems contain interfaces with ferromagnetic films, which are still too complex to purify the bulk SHE from possibly entangled interfacial effects such as the spin mixing conductance, spin memory loss and magnetic proximity effects. An elaborately designed AFM system without any ferromagnetic(FM)/AFM interfaces can promisingly provide a clearer platform to investigate the influence of spin fluctuations on the bulk SHE.

Among antiferromagnets, polycrystalline Cr has already shown sizable $\theta_{SH}$ from -0.051[27] to -0.09[28, 29] and low resistivity as a candidate SOT source. Here, we furthermore choose the epitaxial Cr single crystalline films sandwiched by MgO as the AFM material to uncover the influence of spin fluctuations on the SHE near its $T_N$. To target the goal, specifically, we measured the spin Hall tunneling spectroscopy (SHTS) for the Cr/MgO/Fe fully epitaxial magnetic tunnel junctions (MTJs) which ideally have no direct FM/AFM interface as desired. In this work, we demonstrated a large $\theta_{SH}$ of -0.34 at $T$=200 K by measuring the SHTS for the Cr/MgO/Fe MTJs. The temperature dependence of $\theta_{SH}$ clearly showed an enhancement by 240% near $T_N$=206 K than at 300 K. Furthermore, the spin Hall resistivity followed a clear power law with $T$ in consistent with modeled ones. All these results demonstrate the antiferromagnetic spin fluctuation mechanism offers an effective way indeed to enhance the bulk SHE as $T$ approaching $T_N$.

**AFM Spin Fluctuation**

Another uniqueness of the FSHE is as following: it is attributed to the combinations of several individual scattering processes between conduction electrons and local spins as explained below, instead of a single process accounting for the SHE. These involved local spins are needed to stay correlated within a certain correlation length, which thus makes the FSHE process temperature-sensitive. In theory, taking spin fluctuation near the Curie temperature ($T_C$) into account, Kondo first developed a theory to explain the anomalous Hall effect in ferromagnets[30]. It is further generalized to account for the SHE and ISHE by considering the short-range spin-spin interaction[31] and the long-range dynamical correlation among localized moments[32]. Okamoto *et al.*[32] derived spin Hall conductivity $\sigma_{SH}^{sj}$ and $\sigma_{SH}^{ss}$ due to the extrinsic side-jump and skew-scattering mechanisms, respectively, based on the Kondo's Model, which predicts the nontrivial

SHA enhancement around the ferromagneic quantum critical point due to the critical spin fluctuations. For the AFM spin fluctuation, while the individual scattering processes $\mathcal{F}_{0,1}$, $\mathcal{F}_2$ and $\mathcal{F}_3$ (defined in Ref. 32 and schematically shown in Fig. 1) remain the same as the ferromagnetic case, the microscopic mechanisms of the side-jump and skew-scattering contributions (Fig.1b) to the SHE by the AFM spin fluctuation require further developments (as shown in Methods). Schematically, the microscopic details of the side-jump and skew-scattering mechanisms (Fig.1b) consist of three scattering processes: the $\mathcal{F}_{0,1}$ process contributes 180º back-scattering of antiparallelly polarized conduction electrons (yellow arrow) with local spins (green arrow); the $\mathcal{F}_2$ and $\mathcal{F}_3$ processes result in transverse scattering of electrons depending on the polarization of local spins and scattered electron spins, respectively[32]. Thus, the sequential scattering due to the $\mathcal{F}_{0,1}$ and $\mathcal{F}_2$ ($\mathcal{F}_3$) processes gives rise to the transverse spin current, leading to the side-jump (skew-scattering) mechanism. Temperature is involved in these extrinsic SHE processes by two ways: (1) the more dynamical spins, functioning as the local scattering centers, are activated with the increase in $T$ toward the ordering temperature as shown in Fig. 1a, and (2) the correlation length $\xi$ between the dynamically-activated local spins increases as $T$ approaches the ordering temperature from above, both of which favors the sequential happening of the $\mathcal{F}_{0,1}$ and $\mathcal{F}_2$ ($\mathcal{F}_3$) processes in a higher probability. Moreover, there is no limitation for the above microscopic processes to take place above or below the ordering temperature, in contrast to ferromagnetically ordered systems where the pure SHE is expected to exist only above the transition temperature because below the transition temperature the anomalous Hall effect appears. Therefore, the AFM order is necessary to observe a pure spin-fluctuation-originated contribution to the SHE, i.e., FSHE, without involvement with the anomalous Hall effect. Specifically, one would then expect a clear enhancement of the SHE in an antiferromagnetic system around its Néel temperature at which the correlation length diverges. This hypothesis motivated us to experimentally study the influence of AFM spin fluctuations on the bulk SHE in the clean Cr/MgO/Fe MTJs by the clear SHTS method, which is also instructive to develop the correspondingly theories.

**Single Crystal Cr Junction**

Chromium is a typical antiferromagnet with collinear spin sublattices and a simple body-centered cubic (bcc) structure. $T_N$ of bulk Cr is 311 K and lowerable by reducing thickness. Epitaxial stacks Cr($t$)/MgO(2.3)/Fe(10)/Au(5 nm) ($t$ =7, 10, 25, 50 nm) was deposited on a MgO (001) substrate by the molecular beam epitaxial growth(Methods). Note that Cr in these MTJs were spaced by MgO from Fe, which avoided direct Cr/Fe interfaces and ideally disentangled the targeted bulk Cr SHE from any interfacial spin-mixing, spin memory loss or magnetic proximity effects[33, 34].

The Fe layer was used as a spin-polarizer of charge current and the Cr layer functioned as a spin-analyzer to convert a spin current into a transverse voltage via the ISHE. The high-resolution cross-sectional transmission electron microscope (HRXTEM) was used to check the MTJ structure (Fig. 2a). The low-mag image validated uniform thickness for each layer well corresponding to the nominal one. The black field image (inset of Fig. 2a) allows us to measure the lattice constant of each layer. The lattice constants of the bcc Cr and Fe near interfaces were $a_{Cr}$=2.91 Å and $a_{Fe}$=2.88 Å, close to their bulk values 2.88 Å and 2.87 Å [35]. $A_{MgO}$=4.24 Å [36] was near $\sqrt{2}\ a_{Cr}$=4.11 Å and $\sqrt{2}\ a_{Fe}$=4.06 Å, which implied the epitaxial relationship: $(001)_{MgO}//(001)_{Cr/Fe}$ and $[100]_{MgO}//[110]_{Cr/Fe}$[36]. The electron energy loss spectroscopy (EELS) was performed to map the elementary distributions (Fig. 2b). There was negligible interdiffusion for the Cr, Fe, Mg and Au elements between neighboring layers. Fig. 2c and its inset show the magnetic hysteresis loops as the magnetic field **H** applied along the in-plane $[110]_{MgO}$, $[100]_{MgO}$ and out-of-plane $[001]_{MgO}$ directions by vibrating sample magnetometer (VSM). The loops verified the in-plane easy axis indeed along the $[110]_{MgO}$ direction. This epitaxial relation attributed to the born magnetic easy-axis of Fe and the AFM ordering of Cr along the $[110]_{MgO}$ direction owing to the magnetocrystalline anisotropy. A single 10-nm Cr was deposited on MgO (001) as a control sample and fabricated into the 4-Probe Bar to determine the $T_N$ of Cr. As shown in Fig. 2d, the resistivity $\rho$ gained an extra enhancement due to the disorder-induced scattering via spin fluctuation near $T_N$[36, 37]. The differential resistivity $\rho$ respective to $T$, i.e., ($d\rho/dT$) captured this enhancement clearly. As $T$>206 K, $d\rho/dT$ maintained stable platform with a small slope. From 206 K to lower $T$, $d\rho/dT$ began to acquire an extra slope because spin fluctuation scattering was switched on in this case. With further lowering $T$ and freezing the AFM structure, after a peak at 150 K, $d\rho/dT$ finally approached to a smaller but stable value due to the decrease in magnon density. This behavior, similar with Ref. 37, indicated $T_N$ of Cr was ~206 K. This value is near $T_N$=235 K of the epitaxial 50 nm Cr as reported[37].

**Spin Hall Tunneling Spectroscopy**

To investigate the SHE in Cr, the SHTS and the harmonic lock-in setup[33, 34] were adopted. The raw film was fabricated into 6×6 μm² MTJs with one top electrode and three bottom ones (Fig. 3a). For detecting the ISHE, an ac current $J$ was injected between Electrodes 1 and 3 (Fig. 3b inset). The magnetization $M_{Fe}$ stayed along MgO [110]. Thus the injected current was spin-polarized along [110] before electrons tunneled through the MgO barrier and entered into Cr. SHE in Cr led to the transverse scattering of the tunneling spin current along the MgO [-110] direction so that an electric field between Electrode 2 and 4 was built as $V_{ISHE} \propto \theta_{SH}(\mathbf{j}_s \times \mathbf{s})$. $\mathbf{j}_s(\mathbf{s})$ indicates the flowing direction of the spin current along the film normal (spin-polarized direction determined by the magnetization of Fe, i.e., $\mathbf{M}_{Fe}$). Then, one could control the output voltage $V_{ISHE}$ or resistance ($dV_{ISHE}/dJ$) by $\mathbf{M}_{Fe}$ with an external field $\mathbf{H}$. When $\mathbf{H}$ is applied out-of-plane along the film normal, $dV_{ISHE}/dJ$ vanishes because of the zero $(\mathbf{j}_s \times \mathbf{s})$ term (Methods). When $\mathbf{H}$ was in-plane and parallel or antiparallel to $[110]_{MgO}$, $(\mathbf{j}_s \times \mathbf{s})$ as well as $(dV_{ISHE}/dJ)$ reached its positive or negative maximum. The maximized $|dV_{ISHE}/dJ|$ value was defined as $R_{ISHE}$. When the $\mathbf{H}$ direction fixed, $R_{ISHE}$ measured in Cr had the same (opposite) sign as in Ta (Pt), indicating Cr shared the same $\theta_{SH}$ sign with Ta. This observation accorded with the previous reports[27, 28, 33]. Besides, we also measured SHTS with increasing $J$ from 10 μA to 70 μA. The magnitude of $R_{ISHE}$ keeps nearly the same as expected (Methods).

**Temperature Dependence**

To observe the critical spin fluctuation enhancement, we performed the SHTS

measurement at different $T$ from 50 K to 300 K with an interval of 25 K. The extracted $R_{ISHE}$ values were plotted in Fig. 3c. Below the critical point 206 K, $R_{ISHE}$ increased rapidly as elevating $T$. Above 206 K, $R_{ISHE}$ was reduced as increasing $T$ further. $T_{max}$ where $R_{ISHE}$ was maximized was almost identical to $T_N$. This enhancement behavior in the temperature dependence is also observed in other samples (Extended Data Fig. 4). To eliminate any possible contributions from the spin transport artifacts in Fe, we also adopted the direct spin Hall effect (DSHE)[34] setup for the same device (Fig. 3b lower panel). In the DSHE measurement, a current was applied between Electrode 2 and 4, which produced a spin accumulation at the Cr/MgO interface. A DSHE voltage $V_{DSHE}$ was then detected by the Fe electrode because of the spin accumulation at the Cr/MgO interface. The magnitude of $V_{DSHE}$ is then proportional to $\mathbf{M}_{Fe}$[38, 39]. $V_{DSHE}$ and thus $R_{DSHE} = (dV_{DSHE}/dJ)_{max}$ across the MTJ was collected between Electrode 1 and 3 to evaluate the SHE (Fig. 3b inset). In this setup, no current flowed through the Fe layer, so $V_{DSHE}$ had no chances to be involved with any magnetotransport artifacts in Fe. The $T$-dependence of $R_{DSHE}$ also showed an enhancement around 200 K, coinciding with the ISHE result. Both $T$-dependence indicated an enhancement of SHE near $T_N$ of Cr, which was attributed to the enhanced fluctuation of local spins near $T_N$ as quantitatively analyzed below.

For the same spin polarizer and setup, $(dV_{ISHE}/dJ)$ depends on intrinsic properties of the spin analyzer material, which enables the SHTS to characterize spin Hall materials[26]. To qualify the enhancement, we calculated $\theta_{SH}$ at different $T$. $\theta_{SH}$ can be deduced from Equation (1) given by Liu et al.[34].

$$R_{ISHE} = \left(\frac{dV_{ISHE}}{dJ}\right)_{max} = \frac{\theta_{SH}P\rho}{w} \cdot \frac{\lambda_s}{t} \cdot tanh(t/2\lambda_s) \quad (1)$$

where $P$ denotes the spin polarization of FM layer, $\lambda_s$ is the spin diffusion length of Cr, $t$ and $w$ is the thickness and channel width of Cr layer, respectively. Thus, given the $R_{ISHE}$ values and other parameters, one can estimate $\theta_{SH}$ of the Cr layer. To estimate $\theta_{SH}$, $\lambda_s$ is necessary. As the reported $\lambda_s$ varies dramatically from 2.1 nm to 13.3 nm[27, 28, 40] in literatures, we determined the $\lambda_s$ in our own MTJs with different Cr thicknesses as shown in Fig. 4a. Reforming Equation (1), Equation (2) containing $\lambda_s$ as a fitting parameter could be derived.

$$wtR_{ISHE}/(P\rho) = \theta_{SH}\lambda_s tanh(t/2\lambda_s) = A(t) \quad (2)$$

We used the parameters at 300 K to figure out $\lambda_s$=8.96±0.37 nm (Fig. 4b). Note that the reported $\lambda_s$ of materials with large SHA slightly but monotonously declines[41, 42] with increasing $T$. This factor cannot increase the injected spin current and has no way to contribute an enhanced SHE as $T$ was elevated from low. The enhanced $R_{ISHE}$ and $R_{DSHE}$ signals can be only ascribed to an enlarged SHA directly instead of the variation of $\lambda_s$. Thus we supposed a temperature-insensitive $\lambda_s$ in the fitting of Fig. 4b accordingly. With this $\lambda_s$, $\theta_{SH}$ could be calculated from the equation (1). For instance, $R_{ISHE}$ =3.13 mΩ at 200 K, $P$=0.74 for single crystal Fe/MgO electrode[43], $w$=10 μm, $t$=10 nm, $\rho$=33.89 μΩ·cm, then we have $|\theta_{SH}|$=0.34. The $\theta_{SH}$ in 10 nm Cr equals to -0.34 at 200 K and -0.10 at 300 K (Fig. 4c). The 300 K value is comparable to but higher than the literature values of -0.051 to -0.09[29], probably due to no magnetic interface influences here. Moreover, spin fluctuation near $T_N$ caused a remarkable SHA enhancement by 240% at 200 K compared to 300 K. The 200 K value has already been comparable to that of heavy metals[44], inferring Cr can function as an efficient SOT source. Besides, the Cr film has relatively low resistivity of about 30 μΩ·cm, nearly one order of magnitude lower than ~190 μΩ·cm of the β-phase Ta[14] and 100-300 μΩ·cm of the β-phase W[19] systems, which can further improve its energy-efficiency. These merits, the large SHA and low resistivity, persist below or around the Néel temperature, which facilitates the use of Cr as a SOT channel material in spin-orbitronics.

**Discussion**

To examine the mechanism of the observed SHE, the scaling relation between spin conductivity $\sigma_{SH}$ and conductivity $\sigma$ of Cr is plotted in Fig.4d. The scaling relation does not simply follow $\sigma_{SH} \propto \text{constant}$ for the side-jump or intrinsic mechanism or $\sigma_{SH} \propto \sigma$ for the skew-scattering mechanism. A clear transition point emerges around $T_N$ rather than a gradual crossover behavior due to mechanism switch[45].

Equation (1) showed the main contributing factor to the enhanced $R_{ISHE}$ was $\theta_{SH}\rho$, the spin Hall resistivity. Normally both spin polarization $P$ [46] and the term $\frac{\lambda_s}{t} \cdot \tanh(t/2\lambda_s)$ involving $\lambda_s$ [41, 42] should maintain stable or decrease slightly but monotonically as $T$ increases. Consequently, these two terms could not account for the rising $R_{ISHE}$ before $T_N$. In Fig. 4e we show the $T$-dependence of $|\theta_{SH}\rho|$. When the skew scattering provides dominant contribution to the SHE, $|\theta_{SH}\rho|=|\sigma_{SH}/\sigma^2|$ serves as an indicator of the SOC strength suggested in Ref. 32. $|\theta_{SH}\rho|$ followed a $T^{3.65}$ power law at low $T<T_{\max}$ (Extended Data Fig. 5b). It appears similar to the predicted behavior near the ferromagnetic quantum critical point with $T^{10/3}$. AFM materials could have different power laws, in particular when magnetic ordering appears at finite temperature.

Note that even though the scattering processes of $\mathcal{F}_{0,1}$ and $\mathcal{F}_2$ ($\mathcal{F}_3$) also take place in AFM materials, the way how they cooperate together to contribute to the SHE can be different from the ferromagnetic counterpart. The long-range dynamical correlation length among localized spins which determines the probability of sequential occurrence of the $\mathcal{F}_{0,1}$ and $\mathcal{F}_2$ ($\mathcal{F}_3$) processes can differ between AFM and FM materials. These factors jointly result in a different $T$-power law in Cr from the 10/3 value due to the ferromagnetic critical fluctuation, which can be used as a benchmark and worth further theoretical investigation. When considering AFM spin fluctuation, Kondo model could be generalized to give $\sigma_{SH}^{ss}/\sigma^2 \propto T^6$ for the skew-scattering mechanism and $\sigma_{SH}^{sj}/\sigma^2 \propto T^2 + \alpha T^4$ for the side-jump mechanism, where $\alpha$ is a constant deduced from the $T$-dependence of $\rho$ (Methods). Both expressions could give a good fit to the experimental data, as shown in Fig. 4d, which persuades us to attribute the enhanced SHA to the AFM spin fluctuation. Although the fitting according to the side-jump mechanism gives a higher confidence, it is still hard to claim which mechanism dominates the enhancement, which will be further investigated hereafter.

Above $T_N$, the AFM texture of Cr is no longer stable. On the other hand, the electron-phonon and electron-electron interactions that weakly depend on spins gradually dominate the $T$-dependence of momentum relaxation process with a $T^5$ law[47, 48] and $T^2$ law[49], respectively. These spin-irrelevant processes contribute more to momentum-relaxation scatterings than the spin fluctuation mechanism. As a natural result, $\theta_{SH}\rho$ inclined to decrease gradually.

## Conclusion

In conclusion, we investigated the spin Hall angles and its temperature dependence of Cr in Cr/MgO/Fe fully epitaxial MTJs. Critical spin fluctuation enhancement of $\theta_{SH}$ to -0.34 in Cr by 240% near $T_N$=206 K than 300 K was observed. The dependence of $\theta_{SH}\rho$ on $T$ before $T_N$ was experimentally measured and fitted with the modeled $T$-dependence of the AFM spin fluctuation. This effective mechanism of increasing spin Hall angles can be instructive to design antiferromagnets with much larger spin Hall angles and low resistivity, and further advance AFM applications in SOT devices.

**Methods**

**Sample Preparation**

The single-crystal stacks are prepared in molecular beam epitaxial system with ultrahigh vacuum. Before deposition, the MgO(001) substrate was annealed at 700 °C for one hour to degas the surface, followed by depositing a 10 nm MgO seed layer at 450 °C to block the C impurity diffusion. The Cr layer was then deposited at 31 °C and followed by an in-situ post-annealing at 600 °C to obtain a flat Cr (001) surface. The temperature was maintained at 78 °C and 57 °C to deposit the MgO and Fe layers. A second post-annealing at 480 °C was performed to improve the crystalline quality of Fe. In the end, a 5 nm Au capping layer was deposited at 88 °C to prevent the surface oxidation. Surface structures of each stack were monitored by reflected high-energy electron diffraction (RHEED) throughout the growth process (Extended Data Fig. 1). After magnetism characterization with vibrating sample magnetometer (VSM), the sample is fabricated into tunnel junctions with standard ultraviolet lithography and ion-beam etching process. The junction is surrounded by oxide $SiO_2$ or $Al_2O_3$ to isolating top electrode and bottom electrodes.

**Electrical measurement**

For ISHE and DSHE measurement, ac current with low frequency is applied with Keithley 6221 and voltage is detected with lock-in amplifier SR830. The first harmonic and second harmonic signals are picked-up by two SR830 simultaneously. We use the frequency of 8.3 Hz and a preamplifier SR560 for good signal-to-noise ratio. The ISHE or DSHE signal in the manuscript is only first harmonic signals, which is disentangled with thermoelectric transport effects. The out-of-plane field gives an ordinary Hall signal, which changes linearly with the field magnitude (Extended Data Fig. 2). And the variation of magnitude of the small ac current we use between 10 µA to 70 µA does not affect the $R_{ISHE(DSHE)}$ value and could modified the signal-to-noise ratio (Extended Data Fig. 3). Magnetic field and temperature conditions are offered by QD PPMS. Samples are mounted on a sample holder with Al wire-bounding.

**Theoretical consideration on the SHE by AFM spin fluctuation**

Here, we briefly discuss how the formalisms derived in Ref. 32 for the FM spin fluctuation should be modified when the AFM spin fluctuation is considered. The main difference from the FM fluctuation is that electrons at the Fermi surface are scattered by the AFM fluctuation with its spectral function given by

$$B_{\mathbf{q}}(\omega) = \frac{1}{\pi} \frac{\omega/\Gamma}{\{\delta + A(\mathbf{q}-\mathbf{Q})^2\}^2 + (\omega/\Gamma)^2}.$$

Here, $\delta$ measures the distance from the magnetic ordering as $\delta \propto T - T_N$ at $T > T_N$ and $\delta \propto M^2$ $T < T_N$ with $M$ being the ordered moment, which behaves as $M \propto \sqrt{T_N - T}$ near $T_N$. A constant $A$ is introduced so that $A(\mathbf{q}-\mathbf{Q})^2$ has the unit of energy, and $\Gamma$ is the Landau damping. In contrast to the similar expression in Ref. 32, (i) the momentum dependence is given by $A(\mathbf{q}-\mathbf{Q})^2$ with $\mathbf{Q}$ the magnetic wave vector characterizing the magnetic ordering, and (ii) the damping term $\Gamma$ is independent of momentum. Because of (i), electrons that contribute to the SHE have to satisfy the nesting condition, i.e., momenta $\mathbf{k}$ and $\mathbf{k}+\mathbf{Q}$ have to be on the Fermi surface. This difference may lead to the following modified forms:

$$\sigma_{SH}^{ss} \propto \tau^2 \, \delta \, I^2(T,\delta)$$

for the skew scattering contribution, and

$$\sigma_{SH}^{sj} \propto \tau \, \tilde{I}(T,\delta)$$

for the side jump contribution. As in Ref. 32, $I(T,\delta)$ is given by

$$I(T,\delta) = \frac{1}{(2\pi)^3} \int d^3p \, \frac{1}{\sinh(\hbar \mathbf{v}_F \cdot \mathbf{p}/T)} \frac{\hbar \mathbf{v}_F \cdot \mathbf{p}/\Gamma}{(\delta + Ap^2)^2 + (\hbar \mathbf{v}_F \cdot \mathbf{p}/\Gamma)^2},$$

where the momentum integral variable is changed from $\mathbf{q}-\mathbf{Q}$ to $\mathbf{p}$ by measuring it from the magnetic wave vector $\mathbf{Q}$, and $\mathbf{v}_F$ is the Fermi velocity. $\tau$ is the carrier lifetime, which is assumed to be independent of momentum that satisfies the nesting condition. $\tilde{I}(T,\delta)$ appearing in the expression of $\sigma_{SH}^{sj}$ is given by

$$\tilde{I}(T,\delta) = \frac{1}{(2\pi)^3} \int d^3p \int d\omega \, \frac{1}{\sinh(\omega/T)} \frac{\omega/\Gamma}{(\delta + Ap^2)^2 + (\omega/\Gamma)^2}.$$

A similar expression is derived for the electron self-energy by the AFM spin fluctuation in Ref. 50 (Wölfle and Ziman Phys. Rev. B **104**, 054441 (2021)). Except for the vicinity of the magnetic transition temperature $T_N$, $I(T,\delta)$ and $\tilde{I}(T,\delta)$ behave as $I(T,\delta) \approx T^3/\delta$ and $\tilde{I}(T,\delta) \approx T^2/\sqrt{\delta}$, respectively. Thus, $\sigma_{SH}^{ss}$ and $\sigma_{SH}^{sj}$ behave as

$$\sigma_{SH}^{ss} \propto \tau^2 \, T^6/\delta$$

and

$$\sigma_{SH}^{sj} \propto \tau \, T^2/\sqrt{\delta},$$

respectively. As shown in Fig. 2d, the resistivity shows a smooth crossover behavior from $\rho \sim const. + \alpha T^2$, where $\alpha$ is a constant, at low temperatures (Extended Data Fig. 5a) to $\rho \propto T$ at high temperatures. Thus, the carrier lifetime is dominated by the impurity or disorder and the electron-electron interaction at low temperatures and by the phonon scattering at high temperatures. Assuming that $\delta$ remains constant away from $T_N$ and $\tau \propto \sigma \sim 1/\rho$, one finds

$$\sigma_{SH}^{ss}/\sigma^2 \propto T^6$$

and

$$\sigma_{SH}^{sj}/\sigma^2 \propto T^2 + \alpha T^4.$$

**Acknowledgements:**

This work was supported by the National Key Research and Development Program of China (MOST) (Grant No. 2021YFB3601302), the National Natural Science Foundation of China (NSFC) (Grant Nos. 51831012, 51620105004, and 11974398), the Strategic Priority Research Program (B) of Chinese Academy of Sciences (CAS) (Grant Nos. XDB33000000). C. H. Wan appreciates financial support from the Youth Innovation Promotion Association, CAS (Grant No. 2020008). The research by S.O. was supported by the U.S. Department of Energy, Office of Science, Basic Energy Sciences, Materials Sciences and Engineering Division. N.N. was supported by JST CREST Grant Number JPMJCR1874, Japan, and JSPS KAKENHI Grant number 18H03676. Y. Lu acknowledges the support of the joint French National Research Agency (ANR)-National Natural Science Foundation of China (NSFC) SISTER Project (Grants No. ANR-11-IS10-0001 and No. NNSFC 61161130527), ANR FEOrgSpin project (Grant No. ANR-18-CE24-0017), ANR SIZMO2D project (Grant No. ANR-19-CE24-0005) and ICEEL SHATIPN projects. The sample growth was performed using equipment from the platform TUBE-Davm funded by FEDER (EU), ANR, the Region

Lorraine and Grand Nancy.

**Author contributions**

C. F. and C. W. conceived and designed the experiment. S. O. and N. N. provided the theoretical analysis. Y. L., X. Z., T. M. and J. Q. grew the single crystal films. C. F., X. W., C. G. and J. D. carried out the VSM and TEM characterization. C. F. and C. W. fabricated the devices and conducted the electrical measurement. G. Y., Z. W., N. T., S. S. P. P. and N. N. gave suggestions on the experiments. Y. L., X. H. supervised this study. All authors discussed the results and prepared the manuscript.

**Competing interests**

The authors declare no competing interests.

**Figure**

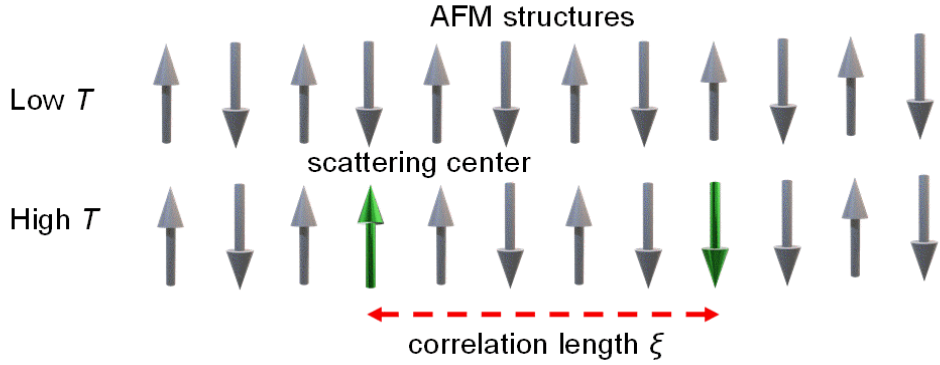

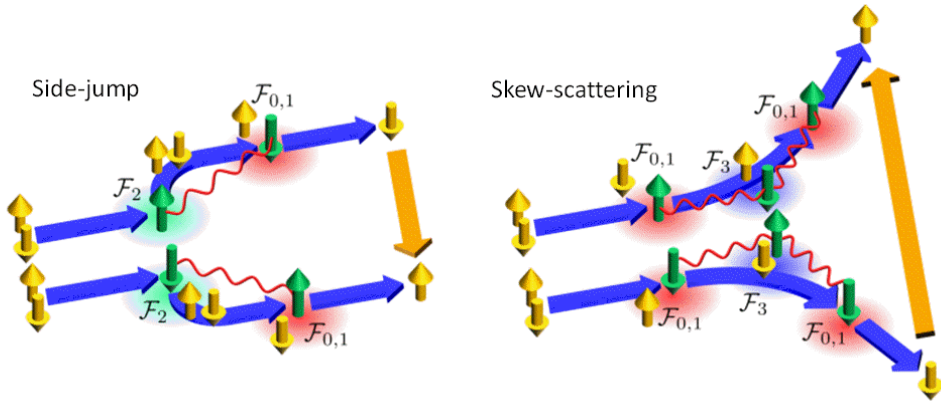

**Fig. 1 | Fluctuation Spin Hall effect. a,** Correlation length $\xi$ of antiferromagnetic materials. The grey arrows are the magnetic spins forming the AFM order. Green arrows act as the scattering centers activated by spin fluctuation. The $\xi$ represents the correlation length within which the scattering centers can antiparallelly correlate with each other. The $\xi$ becomes divergent with $T$ approaching $T_N$ from low temperatures. **b,** Antiferromagnetic spin fluctuation. Spin current (orange allows) is caused by AFM spin fluctuation. Yellow arrows are conduction electron spins, and green arrows are local spins. In the $\mathcal{F}_{2(3)}$ scattering processes, the deflected direction of scattered electrons depends on the direction of the local spins (the conductive electron spins), combining $\mathcal{F}_{0,1}$ to form the side-jump-type (skew-scattering-type) contribution to $\sigma_{SH}$. The red wavy curves represent the propagators of AFM spin fluctuation, which relates to the correlation length.

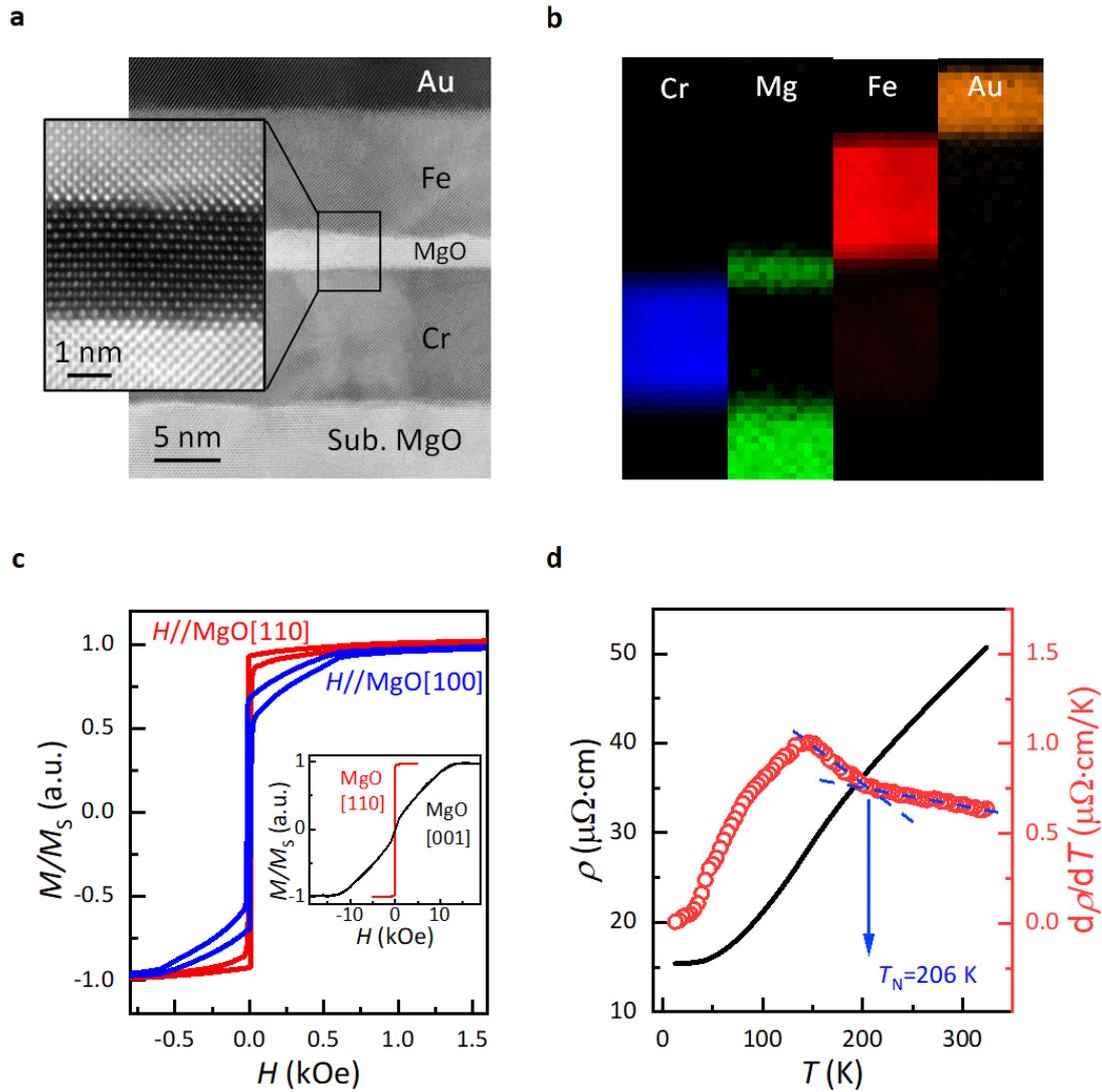

**Fig. 2 | Magnetic tunnel junction Cr/MgO/Fe/Au stacks. a,** High resolution cross-sectional transmission electron microscope (HRXTEM) pattern of the junction. Flat interfaces are indicated. Inset: Black field image of the region near MgO barrier. Epitaxy relation was well defined globally throughout the sample. Cr and Fe lattices grew along the MgO [001] direction. **b,** The electron energy loss spectroscopic (EELS) mapping of the Cr, Fe, Mg and Au elements. **c,** Magnetic hysteresis loop of MTJ films with external magnetic field along MgO [110] (red, also noted as in-plane) and MgO [100] (blue). Inset: MgO [001] (black, also noted as out-of-plane) and reproduction of MgO [110] to show the in-plane easy axis of Fe layer. **d,** The temperature dependence of resistivity (black line) and its differential with respect of $T$ (red circles) in 10 nm Cr.

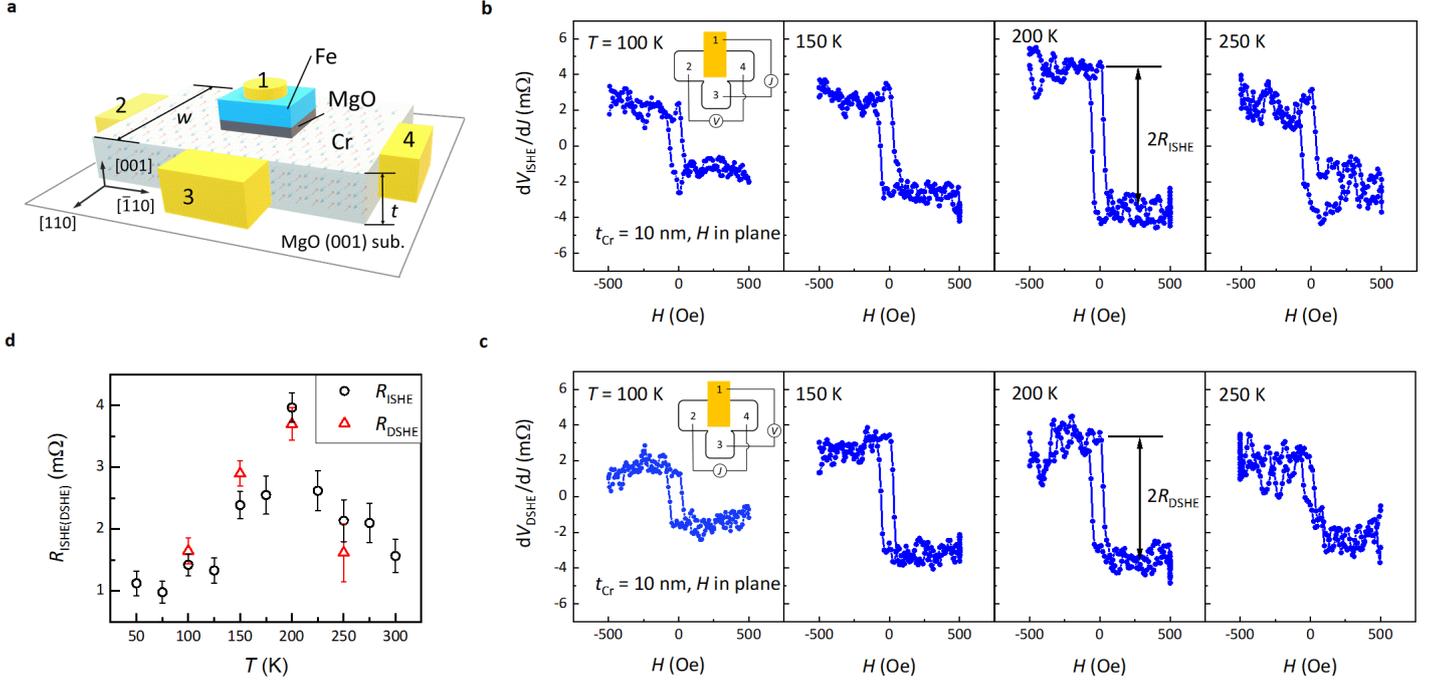

**Fig. 3 | Detection of SFHE with ISHE and DSHE set-up. a,** Device schematic diagram of fabricated MTJ devices. **b,** ISHE and **c,** DSHE in 10 nm Cr with in-plane magnetic field along MgO [110] at different temperatures. Inset: Measurement schematic diagram of ISHE with current $J$ loaded between electrode 1 and 3 and $V_{ISHE}$ collected between electrode 2 and 4 and DSHE with exchanging the source and measure meters. Both ISHE and resistance was observed when **H** was applied in in-plane (blue) and vanished when **H** was out-of-plane applied (Extended Data Fig. 2). $R_{ISHE(DSHE)}$ represented the saturated value of $dV_{ISHE(DSHE)}/dJ$. **d,** Critical spin fluctuation enhanced $R_{ISHE}$ (black circles) and $R_{DSHE}$ (red squares) resistance in MTJ device. The magnitude of $2R_{ISHE(DSHE)}$ is given by the difference between intercepts of linear fitting of two resistance platform and the error bar is given by the standard error of the linear fitting.

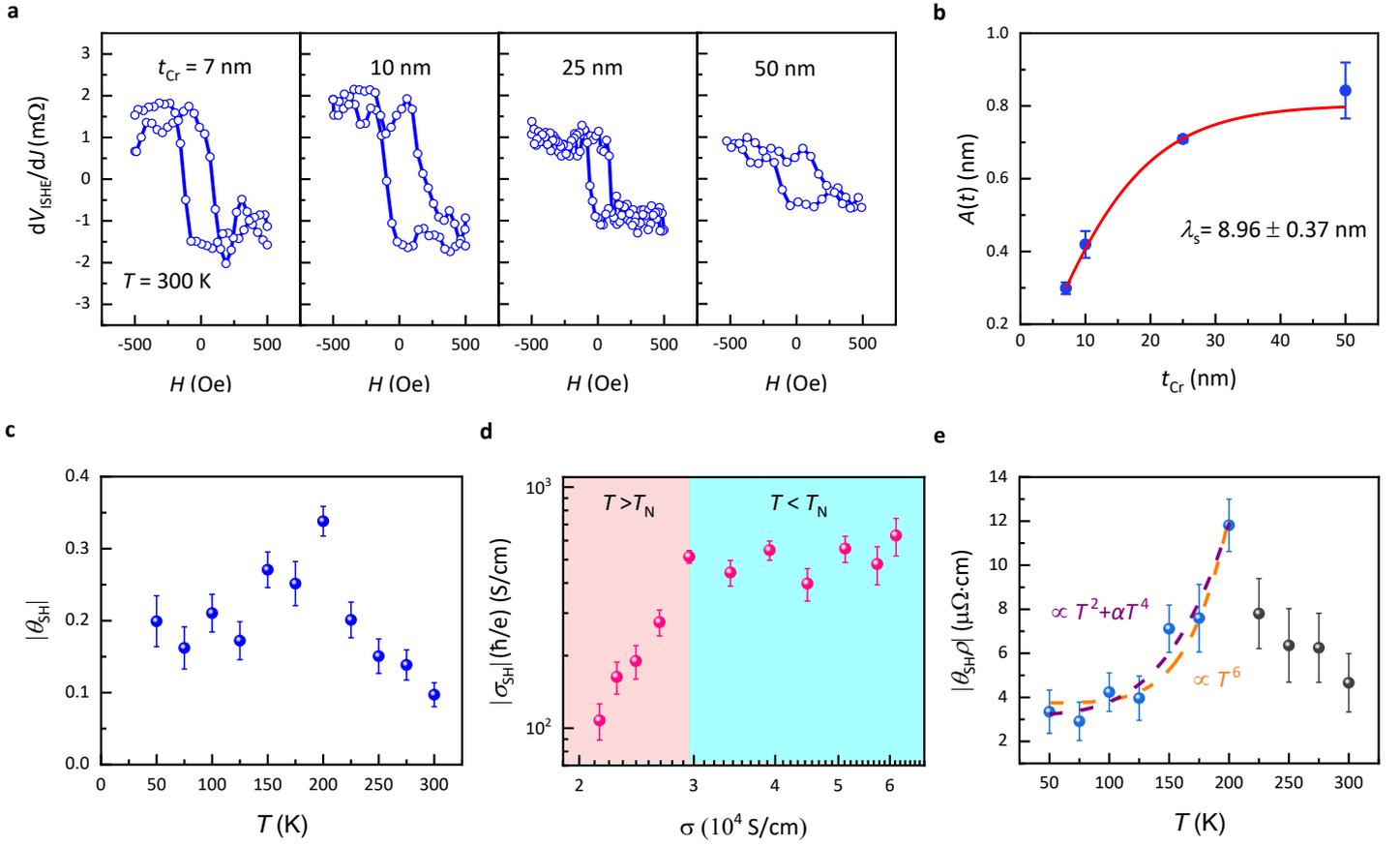

**Fig. 4 | Spin transport properties in Cr. a,** ISHE resistance in Cr at 300 K with different thicknesses. **b,** Thickness dependence of the factor $A(t)$. Red line is the fitting result following Equation (2). **c,** Temperature dependence of SHA. **d,** Scaling relation between conductivity and spin Hall conductivity. **e,** Temperature dependence of product of SHA and resistivity. Orange dash line shows the fitting result of $\sigma_{SH}^{sj}/\sigma^2 \propto T^2 + \alpha T^4$ with the adjusted $R^2$ (coefficient of determination) = 0.95. Purple dash line shows the fitting result of the $\sigma_{SH}^{ss}/\sigma^2 \propto T^6$ with the adjusted $R^2$ = 0.89.

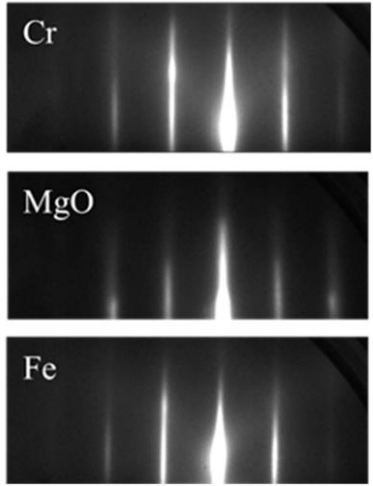

**Extended Data Fig. 1| RHEED pattern of MTJ film.** RHEED pattern of the annealed Cr, MgO and Fe layers further confirmed the epitaxial growth mode. The sharp streaky patterns indicate a high-quality epitaxial growth character of our sample.

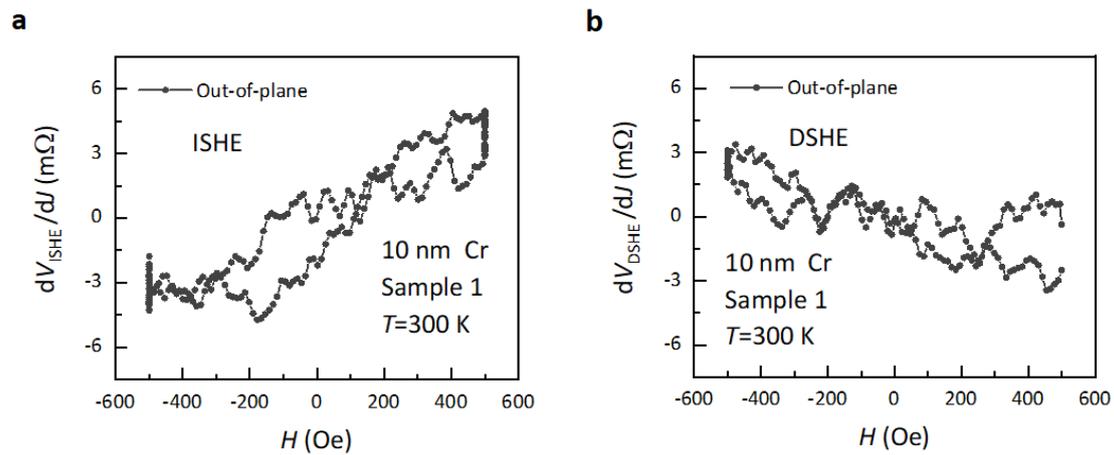

**Extended Data Fig. 2| Spin Hall signal with out-of-plane magnetic field *H*. a,** ISHE. **b,** DSHE

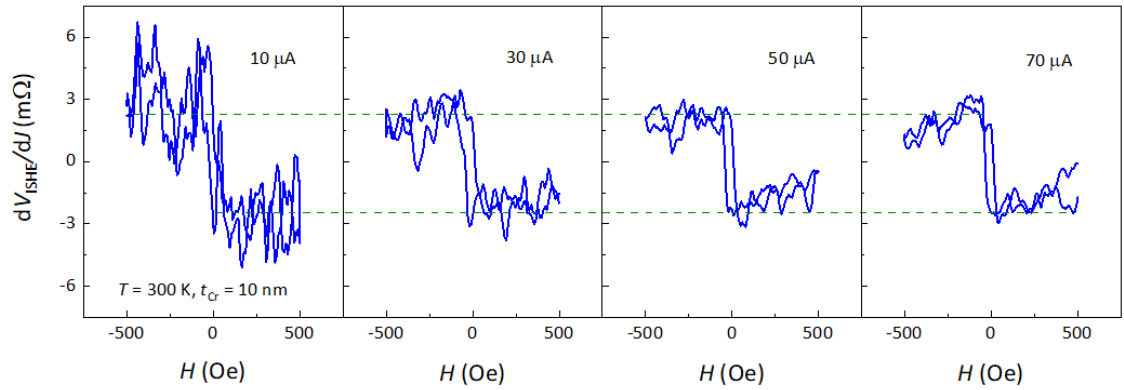

**Extended Data Fig. 3 | Spin Hall signal with different magnitude of injection current**

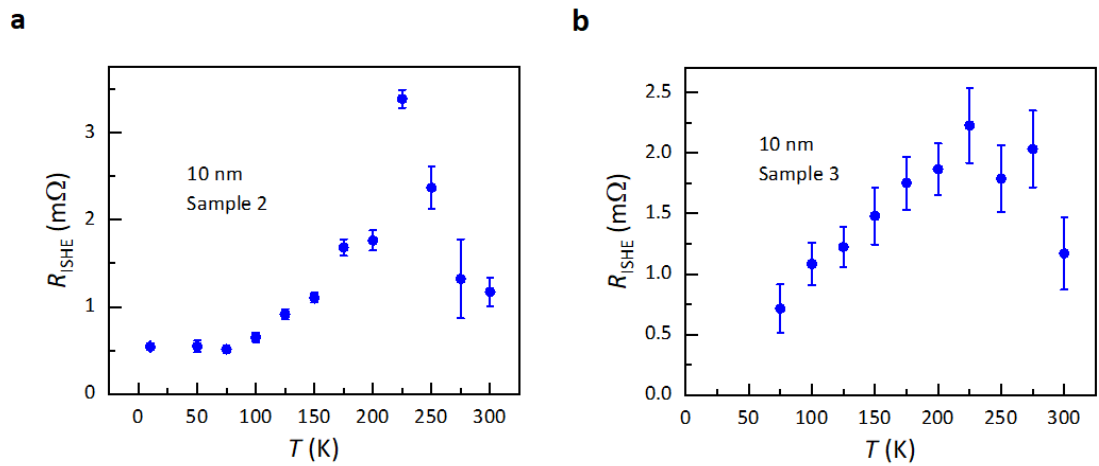

**Extended Data Fig. 4 | Inverse Spin Hall signal in more samples. a,** Sample 2. **b,** Sample 3.

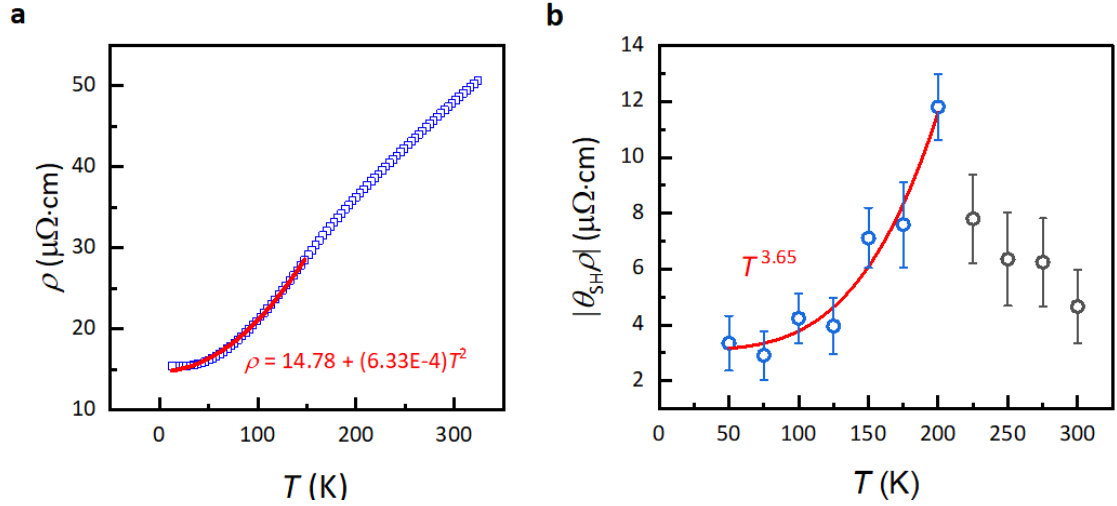

**Extended Data Fig. 5 | Fitting of *T*-dependence of resistivity and spin resistivity. a,** Resistivity fitting. Data points with the square shape are repeat of ones in Fig. 2d. Fitting $\rho \sim const. + \alpha T^2$ gives the $\alpha$. **b,** Spin resistivity fitting. Data points with the circle shape are repeat of ones in Fig. 4e. Red line indicates a power-law fitting of the spin Hall resistivity at and below 200 K.